\documentstyle[preprint,aps,version2]{revtex}
\begin{document}
\draft

\begin{title}
Novel Structure Function for Photon Fragmentation
into a $\Lambda$ Hyperon  and Transverse $\Lambda$ Polarization
in  Unpolarized Electron-Positron Annihilation
\end{title}

\author{Wei Lu}

\begin{instit}
CCAST (World Laboratory), P.O. Box 8730, Beijing 100080, China

and Institute of High Energy Physics,
 P.O. Box 918(4), Beijing 100039, China\footnote{Mailing address}
\end{instit}

\begin{abstract}
The possibility is examined for the inclusive $\Lambda$ in unpolarized
electron-positron annihilation  to be transversely polarized.
Due to  final-state interactions,  there exists a
novel structure function $\hat F(z,Q^2)$ for  the  inclusive $\Lambda$
hyperon (or any other baryons) production from the unpolarized
time-like photon fragmentation, which makes contribution
to the transverse $\Lambda$ polarization
in the unpolarized electron-positron annihilation.
\end{abstract}
\pacs{ PACS numbers: 13.88.+e, 13.65.+i}

\pagestyle{plain}
During the past two decades, quantum chromodynamics (QCD)
has been established gradually as the most  prospective
candidate for the underlying strong-interaction theory.
In interpreting various spin asymmetries and  final-state
transverse polarization phenomena,  however, it is not so
successful. Early in 1970's, for example,  people
\cite{Fermi} at Fermilab observed
large $\Lambda$ polarization perpendicular to the production plane
in unpolarized  p-${\rm B_e}$ experiments.
Until now,  no completely QCD-based  scenario  has
been successfully established to explain such seemingly
spectacular phenomena.

   By the  standard helicity amplitude analysis
\cite{Wick},  the transverse
polarization of inclusive fermion  is  proportional to
the interference of its  two helicity amplitudes. In a $massless$
vector or  axial-vector theory, if there is no  novel physical
mechanism,  there  exists no transverse polarization.
As the  effects of particle masses are included,
intuitively,   the mass-induced  polarization is to
be  proportional to the mass parameter. It is based
on this fact that Kane, Pumplin and Repko \cite{Kane}
pointed out that hard partonic scatterings cannot
take the whole responsibility for the large transverse
$\Lambda$ polarization in hadronic processes.  In general,
the transverse quark polarization due to hard partonic
subprocesses assumes the following asymptotic behavior:
\begin{equation}
P_q\propto  \displaystyle\frac{\alpha_s m_q}{Q},
\end{equation}
where $Q$ is  the scale of hard scatterings.
In the parton model, the parton  fragments into  hadrons
independently, so the transverse  polarization of
inclusive baryon can  only come from  that
of partons responsible for its production.
Since so,  the search is in need  for other
possible resources of the large transverse $\Lambda$ polarization.

 For the inclusive fermion production, if we denote the
momenta of   the beam  and  inclusive  fermion
by ${\bf p}$ and ${\bf k}$ respectively,  the transverse fermion
polarization is  proportional to the expectation  value
$<{\bf s}\cdot ({\bf p} \times {\bf k} )>$.
Despite ${\bf s}\cdot ({\bf p} \times {\bf k} )$ is
time-reversal odd, final-state
interactions may  make contributions
to the transverse polarization \cite{Gasi}.
Therefore, it is an interesting theme to
examine how well final-state
interactions  contribute to the  transverse polarization
of inclusive observed $\Lambda$ hyperons in fixed-target
experiments.   At present, it is not an  easy  job
because of the complexities of  hadronic processes.

Alternatively, we investigate in this Brief Report  the  possibility
of studying the  effects of final-state interactions  in the process
\begin{equation}
\label{main}
e^- (p)  + e^+ (q-p) \to \Lambda (k, s) +X,
\end{equation}
where $s^\mu$ is the $\Lambda$ spin vector and normalized
as $s\cdot s=1$ for a pure state.
Naively,  since the initial-state   particles  are  unpolarized,
the inclusive $\Lambda$  production seems to have no spin weights.
Owing to the interactions between the final-state particles,
however,  there do exist possibilities
for the inclusively detected $\Lambda$'s  to be transversely polarized,
as we will show in the coming discussion.

We choose to work in the energy region far below the $Z$-resonance. In
such a case,  electron-positron  annihilation proceeds predominantly
via  the  single-photon  channel.  Correspondingly, its
differential cross section may be written as
\begin{equation}
{\rm d}\sigma=\displaystyle\frac{1}{2Q}~(\displaystyle\frac{4\pi\alpha}{Q^2})^2
(4\pi)L_{\mu\nu}\tilde W^{\mu\nu}
\displaystyle\frac{{\rm d}^3 k}{(2\pi)^3 2E},
\end{equation}
where  $Q=\sqrt{q^2}$ is the c.m. energy,
 $L_{\mu\nu}$ the lepton tensor,  and $\tilde W_{\mu\nu}$
the photon fragmentation tensor for  the inclusive
$\Lambda$ production. We isolate the electromagnetic
coupling constant $e^2=4 \pi \alpha$
for each tensor in the sense that  the electromagnetic
current is defined as $j_\mu= Q_f\bar \psi \gamma_\mu \psi$.
In addition, the electron mass is ignored.

Now we dwell on the Lorentz decomposition of
photon fragmentation tensor $\tilde W_{\mu\nu}$.
In the case of inclusive baryon production, it is
considered  usually  without spin information.
In this Brief Report, we concentrate ourselves
on the case of measuring  the
spin of inclusively produced $\Lambda$ hyperon.

The photon fragmentation tensor for the
inclusive $\Lambda$ production is defined as
\begin{equation}
\label{tensor}
\tilde W_{\mu\nu}(k,q,s)=\displaystyle\frac{1}{4\pi}
\int {\rm d}^4 x e^{i q\cdot x}
<0|j^\dagger_\mu(0)|\Lambda(k,s),X>_{\rm out}~
_{\rm out}<\Lambda(k,s),X|j_\nu(x)|0>.
\end{equation}
Throughout our work, the summation  over  the unobserved system $X$
is understood.  The general  Lorentz decomposition
of $\tilde W_{\mu\nu}(q,k,s)$
is  subjected   to parity conservation,  time reversal invariance
and current conservation.  In addition,  the
hermiticity of $j_\mu$ leads to
\begin{equation}
\tilde W_{\mu\nu}(k,q,s)=\tilde W^\ast_{\nu\mu}(k,q,s).
\end{equation}
In obtaining the general expression for $\tilde W_{\mu\nu}(q,k,s)$,
the Lorentz vectors and
tensors to be employed
are  $k_\alpha$, $q_\alpha$, $s_{\alpha}$,
$g_{\alpha\beta}$ and $\varepsilon_{\alpha\beta\gamma\delta}$.

Let us imagine a process of
deeply inelastic scattering for the unpolarized
electron beam and polarized $\bar \Lambda$ target.
Then, process (\ref{main}) can be related to it by
reversing the momenta and  spin vector  of
position and $\Lambda$ particle.  Intuitively,
the hadronic tensor for $\bar \Lambda$
and photon fragmentation tensor  for the inclusive
$\Lambda$ production should bear the  same Lorentz
structures, since  both two  are subjected to the same
symmetries.  The hadronic tensor of spin-half
targets  has been well-known \cite{hard}.  By analogy,
the decomposition of $\tilde W_{\mu\nu}(k,q,s)$ contains
at least  the following  combinations:
$$-g_{\mu\nu}+\displaystyle\frac{q_\mu q_\nu}{q^2},~
(k_\mu -\displaystyle\frac{k\cdot q}{q^2}q_\mu)
(k_\nu -\displaystyle\frac{k\cdot q}{q^2}q_\nu),~
i\varepsilon_{\mu\nu\lambda\sigma}q^\lambda s^\sigma,~
i\varepsilon_{\mu\nu\lambda\sigma}q^\lambda
(s^\sigma-\displaystyle\frac{s\cdot q}{k\cdot q} k^\sigma).$$
In the case of unpolarized electron-positron annihilation,
parity  conservation determines that
the inclusive $\Lambda$  can be  transversely polarized only.
Correspondingly, $s \cdot q=0$, and  the last two  of the
combinations displayed above are degenerate.   However,
if initial-state particles are polarized, $s \cdot q$
vanishes no longer. Although our discussions will
 be concentrated on the unpolarized annihilation,
formally we still distinguish  the above two  spin-dependent  terms.

Our findings to be presented are based  upon the  following
observation. That is, the  hadronic
tensor is built on a single-particle state only, while fragmentation
tensor involves  multi-particle final states.  Such a difference
causes their distinct behaviors under time reversal
transformation,  because  under time reversal transformation
an out-state  is changed into its corresponding in-state.
Unless it is a single-particle state, an out-state differs from
its corresponding in-state by a unitary transformation S \cite{Bj}.
Generally, the S  operator (matrix) can be written as
\begin{equation}
|>_{\rm in}=S|>_{\rm out},~~{\rm with}~~ S=1+i T,
\end{equation}
where $T$ is  the transition  operator (matrix).

We examine how  the fragmentation tensor  behaves
under $PT$ transformation instead of individual  transformations
$P$ and $T$. Under $PT$ transformation, spin vector
changes sign while  momentum
does not, correspondingly it is convenient to study spin asymmetries.
In our case,
\begin{equation}\begin{array}{ll}
\label{bonnie}
\tilde W_{\mu\nu}(k,q,s)&=
\displaystyle\frac{1}{4\pi}\int {\rm d}^4 x e^{i q\cdot x}
[<0|j^\dagger_\mu(0)| \Lambda(k,-s),X >_{\rm in}~
 _{\rm  in} <\Lambda(k,-s),X|j_\nu(-x)|0>]^\ast\\
{}~~~~& ~~~~~~~\\
{}~~~~& =[\displaystyle\frac{1}{4\pi}\int {\rm d}^4 x e^{i q\cdot x}
<0|j^\dagger_\mu(0)S|\Lambda(k,-s),X  >_{\rm out}~
 _{\rm  out} <\Lambda(k,-s),X|S^\dagger j_\nu(x)  |0>]^\ast\\
{}~~~~& ~~~~~~~\\
{}~~~~& =[\displaystyle\frac{1}{4\pi}\int {\rm d}^4 x e^{i q\cdot x}
<0|j^\dagger_\mu(0)|\Lambda(k,-s),X  >_{\rm out}~
 _{\rm  out} <\Lambda(k,-s),X| j_\nu(x)  |0>]^\ast
+\cdot\cdot\cdot . \\
{}~~~~& ~~~~~~~\\
{}~~~&= \tilde W_{\mu \nu}^\ast (k,q,-s) +\cdot \cdot \cdot .
\end{array}
\end{equation}
This is completely different from  the case  of deeply
inelastic scattering, in which one can derive $W_{\mu\nu}(q,p,s)
= W_{\mu\nu}^\ast (q,p,-s)$. The terms denoted by ellipses
in Eq. (\ref{bonnie})  are  not subjected to the constraint
$\tilde W_{\mu\nu}(q,p,s)= \tilde W_{\mu\nu}^\ast (q,p,-s)$.
As a consequence, the general expression of $\tilde W_{\mu\nu}$  may
contain  other  terms rather than  those that have counterparts
in $W_{\mu\nu}(q,p,s).$

Physically, these novel terms
are due to the interactions  between final-state
particles [7,8]. To see this,
let us examine the  following  term  that was suppressed
in Eq. (\ref{bonnie}):
$$-\displaystyle\frac{i}{4\pi}[\int {\rm d}^4 x e^{i q\cdot x}
<0|j^\dagger_\mu(0)T|\Lambda(k,-s),X  >_{\rm out}~
 _{\rm  out} <\Lambda(k,-s),X| j_\nu(x)  |0>]^\ast.$$
Making use of the completeness of final  states,  we
can insert $\int {\rm d} \Gamma |{\rm f.s.}>_{\rm out}
{}~_{\rm out}<{\rm f.s.}|=1$  before the transition operator $T$,
where $|{\rm f.s.}>$ and ${\rm d} \Gamma$ stand for the final state
and its phase space elements, and the integration
includes the summation over all the discrete degrees
of  freedom. Then we have the following physical  picture:
A time-like photon decays first into  other
possible intermediate $physical$  states, which
then  transit via interactions  to the  inclusively
detected final state. It is such final-state
scatterings that make  the relative  phase between an in-state and
its corresponding out-state.

Subject  to the constraints from all symmetries,
the novel terms arising from final-state interactions
can only be of the Lorentz  structure
$$[(k_\mu-\displaystyle\frac{k\cdot q}{q^2}q_\mu)
\varepsilon_{\nu \alpha \beta \gamma}
k^\alpha q^\beta s^\gamma
+(k_\nu-\displaystyle\frac{k\cdot q}{q^2} q_\nu)
\varepsilon_{\mu \alpha \beta \gamma}
k^\alpha q^\beta s^\gamma ].  $$
Thus, the full decomposition of photon  fragmentation tensor reads
\begin{equation}\begin{array}{ll}
\label{dec}
\tilde W_{\mu\nu}(k,q,s)&
=\displaystyle\frac{1}{2}
[ (-g_{\mu\nu}+\displaystyle\frac{q_\mu q_\nu}{q^2})\tilde F_1
+(k_\mu -\displaystyle\frac{k\cdot q}{q^2}q_\mu)
(k_\nu -\displaystyle
\frac{k\cdot q}{q^2}q_\nu)\displaystyle\frac{\tilde F_2}{k\cdot q} ]
 \\
{}~~~&~~~\\
{}~~~&
+iM_\Lambda \varepsilon_{\mu\nu\lambda\sigma}q^\lambda s^\sigma
\displaystyle\frac{\tilde g_1}{k\cdot q}
+iM_\Lambda \varepsilon_{\mu\nu\lambda\sigma}q^\lambda
(s^\sigma-\displaystyle\frac{s\cdot q}{k\cdot q} k^\sigma)
\displaystyle\frac{\tilde g_2}
{k\cdot q}\\
{}~~~&~~~\\
{}~~~&
+M_\Lambda [(k_\mu-\displaystyle\frac{q\cdot k}{q^2}
q_\mu)\varepsilon_{\nu \alpha \beta \gamma}
k^\alpha q^\beta s^\gamma
+(k_\nu-\displaystyle\frac{q\cdot k}{q^2}q_\nu)
\varepsilon_{\mu \alpha \beta \gamma}
k^\alpha q^\beta s^\gamma ] \displaystyle\frac{\hat F}{(k\cdot q)^2}.
\end{array}\end{equation}
The factor $\displaystyle \frac{1}{2}$ in the
spin-independent terms  implies the fact that we must
perform spin summation if we do not measure the $\Lambda$
spin. $\hat F$ is a novel structure function
arising from the final-state interactions.
We have borrowed the labels from
the  expression of hadronic tensor  for the other
four structure functions and added  to them a tilde.
$\tilde F_1$ and $\tilde F_2$ are spin independent,
while $\tilde g_1$, $\tilde g_2$ and $\hat F$
are spin depedent. By the  hermiticity of electromagnetic
current, all  five fragmentation structure functions
are real. In addition, all of them  depend
on $z=\frac{2k\cdot q}{q^2}$ and $Q^2=q^2$.

The novel structure function $ \hat F$ is closely related to
the transverse polarization of the inclusive $\Lambda$ hyperon
in the initial-state unpolarized electron-positron annihilation,
 which is defined by
\begin{equation}
P_\Lambda=\displaystyle\frac
{{\rm d}\sigma(s_\uparrow)-{\rm d}\sigma(s_\uparrow)}
{{\rm d}\sigma(s_\downarrow)+{\rm d}\sigma(s_\downarrow)}.
\end{equation}
Recall that when the initial-state electron and positron
are unpolarized, lepton tensor is both real and symmetric
under $\mu \leftrightarrow \nu$.  Therefore, only the
novel term in Eq. (\ref{dec}) contributes to the $\Lambda$-spin
dependence of cross section.  Schematically, we have
\begin{equation}
P_\Lambda=\displaystyle\frac
{ L_{\mu\nu}
\tilde W^{\mu\nu}(q,k,s_\uparrow)|_{\rm novel}
}
{
L_{\mu\nu}\tilde W^{\mu\nu}(q,k,s_\uparrow)|_{\rm spin~independent}
}.
\end{equation}
After a little algebra, we obtain
\begin{equation}
P_\Lambda=
\displaystyle\frac{2M_\Lambda \sin\theta \cos\theta\hat F}
{
Q(1+\displaystyle\frac{M^2_\Lambda}{{\bf k}^2})\tilde F_1
+\displaystyle\frac{1}{2}
\sqrt{{\bf k}^2+M^2_\Lambda}\sin^2\theta \tilde F_2
}, ~~{\rm with}~~ \theta=\angle({\bf p}, {\bf k}).
\end{equation}

The structure functions $\tilde F_1$ and $\tilde F_2$
are  associated with the inclusive $\Lambda$ hyperon production
without  measuring its spin. If one works with
the parton model,   $\tilde F_1$ is related to
$ \tilde F_2$ by $ 2\tilde F_1+z \tilde F_2=0$. At present,  the
$\Lambda$ production study based on  the BES data at BEPC
is under the  way \cite{Xie}.  It can be anticipated
that  the measurements of $\tilde F_1$ and $\tilde F_2$
are realized in the near future.

As we  have identified, $\hat F$ incorporates the effects of
final-state interactions on the transverse $\Lambda$
polarization in unpolarized electron-positron annihilation.
Since it evolves multiple interactions and the  hadronic
parton interactions become more and more  dominant in the
high-energy case,    the effects  of $\hat F$
are  power-suppressed, or so-called higher-twist effects.
 In fact, all transverse spin
asymmetries are  twist-three effects \cite{QS}.
It is known that  in the moderate kinematical region
some higher twist effects  can become relatively large \cite{twist}.
In the energy region in which BEPC,  TRISTAN and the  proposed
Beauty  and Tau-Charm Factories lie,  it is  quite probable
that  final-state interactions  are not  shadowed  by hard
partonic physics. Here, we suggest measuring the transverse polarization
of  the inclusively produced $\Lambda$ hyperon in
the initial-state unpolarized electron-positron annihilation.
Process (\ref{main}) is an ideal place to examine
how well  final-state interactions
contribute to the  transverse $\Lambda$ polarization.
Since  the final-state interactions are widely existent in
any inclusively detected   particle system, we may
anticipate the measurement of  transverse $\Lambda$ polarization in
unpolarized  electron-positron annihilation will
help our understanding of  transverse hyperon polarization phenomena.
Whether  positive or negative, the  results of  the
suggested experiment will be enlightening. On the one hand,
if a  large transverse   polarization is detected, then we
may expect that final-state interactions play a
significant role in the transverse $\Lambda$ polarization
in  fixed-target  processes, although these experiments
are done at somewhat different energy scales.
  On the other hand,  if  there
is  little transverse polarization in process (\ref{main}),  we are
supposed to search for other possible  resources for
the large transverse $\Lambda$ polarization  phenomena.

In conclusion,  we  examined the general Lorentz
decomposition of photon fragmentation tensor for the
inclusive $\Lambda$ hyperon production.  By emphasizing
the difference between the fragmentation tensor and
hadronic tensor in deeply inelastic scattering, we point
out that  there exists a novel term in the  general
expression of the former, which is associated  with
the  possible transverse $\Lambda$ polarization
in unpolarized electron-positron
annihilation. Physically, such a transverse polarization
is due to the  interactions between the final-state
particles. We suggest  measuring the transverse polarization
in process (\ref{main}) so  as to obtain some indications
how well  final-state interactions contribute to the
large transverse $\Lambda$ polarization in fixed-target
experiments.

\acknowledgements{The author is indebted to T. Huang, X.Q. Li,
C.W. Luo and Y.G. Xie for encouraging discussions.}


\begin{thebibliography}{150}
\bibitem{Fermi}
G. Bunce $et$  $al.$, Phys. Rev. Lett. {\bf 36}, 1113 (1976)
\bibitem{Wick}
M. Jacob and G.C. Wick, Ann. Phys. {\bf 7}, 404 (1959)
\bibitem{Kane}
G.L. Kane, J. Pumpkin, and W. Repko, Phys. Rev. Lett.
{\bf 41}, 1689 (1978)
\bibitem{Gasi}
S. Gasiorowicz, $Elementary ~Particle ~Physics$
 (Wiley, New York, 1966).
\bibitem{hard}
B.L. Ioffe, V.A. Khoze and L.N. Lipatov, $Hard ~ Processes$
(Elsivier, Amsterdam, 1984), Vol. 1.
\bibitem{Bj}
J.D. Bjorken and S.D. Drell, $Relativistic ~Quantum ~Fields$
(McGraw-Hill, New York, 1965).
\bibitem{J_J}
R.L. Jaffe and X. Ji, Phys. Rev. Lett. {\bf 71}, 2547 (1993)
\bibitem{Ji}
X. Ji, Phys. Rev. D {\bf 49}, 114 (1994).
\bibitem{Xie}
Y.G. Xie, private communication.
\bibitem{QS}
J. Qiu and G. Sterman, $Polarized ~Collider ~Workshop$,
edited by J.C. Collins, S.F. Heppelman, and R.W. Robinett
(American Institute of Physics, New York, 1990) p. 249;
Nucl. Phys. B {\bf 353}, 137  (1991).
\bibitem{twist}
J. Qiu and G. Sterman, Phys. Rev. Lett. {\bf 67}, 2264 (1991);
Nucl. Phys. B {\bf 378}, 52 (1992).
\end{thebibliography}
\end{document}